\documentclass{aastex}

\begin{document}

\title{COOLING A HOT DISK AROUND A SUPERMASSIVE BLACK HOLE BY A STAR}

\author{Heon-Young Chang}

\affil{Korea Institute For Advanced Study, Seoul, 130-012, Korea}

\email{hyc@kias.re.kr}

\begin{abstract}

If a supermassive black hole resides in the centers of galaxies,
there are several effects expected to be observed. It is likely 
that accretion disks are around the supermassive black hole.
Stellar interactions with the accretion flows around 
the supermassive black hole play a role in that an flying-by star 
may cool a hot accretion disk as a result of Comptonization. It is shown
that the  Comptonization of the stellar emission will take place in 
a hot accretion disk such as the ADAFs around the supermassive black hole
and become a relatively important source of the accretion disk cooling 
when the mass accretion rate is low, and stars are passing 
outer parts of the hot accretion disk. We suggest that such a stellar cooling 
can be observed in the radio frequency regime since synchrotron luminosity
of the ADAFs depends strongly on the electron temperature and occurs
much more frequently than a tidal disruption event.

\end{abstract}


\section{INTRODUCTION}

Since the early days of research on quasars and AGNs, supermassive black holes
(SMBHs) have been considered as the most likely power sources of the activity 
in these objects \citep{lyn69,rees84}. Many lines of direct and indirect evidence 
suggest that SMBHs reside in the centers of a substantial fraction of galaxies 
\citep{kor95}. If this is the case, several effects are expected 
to be observed. When stars approach to the SMBH in the center of galaxies, 
stars can be captured or tidally disrupted by the SMBH in galactic nuclei, 
resulting in flare-like activity. 
Several aspects of disruptive stellar encounters with supermassive 
black holes have been studied both theoretically and observationally 
(e.g., Nolthenius \& Katz 1982; Bicknell \& Gingold 1983; 
Carter \& Luminet 1983; Peterson \& Ferland 1986; Hills 1988; Kochanek 1992; 1994; 
Syer \& Clarke 1992; 1993; Laguna et al. 1993; Brandt et al. 1995; 
Renzini et al. 1995; Storchi-Bergmann et al. 1995; Marck et al. 1996).
During such a dramatic event, 
energy outbursts  are expected in different processes
  and its observational implications are various. 
Explosive release of nuclear energy due 
to the extreme compression for the orbiting star may occur \citep{lum89,lum90} 
and may lead to observable enrichment of heavy elements in
the galactic center. Tidal disruption events are rare, but seemingly 
unavoidable consequences of $10^6 - 10^8 M_{\odot}$ SMBHs in galactic centers.
About half of the stellar debris is bound and accretes onto the central
SMBH, and the rest is flung-out. The bound half may create a bright UV/X-ray flare 
as it accretes onto the SMBHs in a timescale of less than a couple of years \citep{rees88}. 
The detection of a single tidal disruption event could even allow 
for a crude estimate of the SMBH mass (e.g., Loeb \& Ulmer 1997). 
\citet{kho96} show that interactions between the high velocity unbound material
and the interstellar medium (ISM) may result in a supernova-like remnant 
with properties similar to those of Sagittarius A East. In extremely low 
angular momentum encounters, the material could deposit over $10^{52}$ ergs 
into the ISM, creating a shell like a supernova remnant. More typically, 
the energy deposited would be $\sim 10^{50}$ ergs.  
Interaction of orbiting debris may occur after the most tightly bound 
debris begins its another orbit \citep{kpl99}.

If there is an accretion disk around the SMBH, a less dramatic process 
may also occur due to the interactions of a flying-by star and the accretion
disk around the SMBH (e.g., Syer et al. 1991; Hall et al. 1996).
The process has become interesting particularly when the accretion disk is 
geometrically thick and hot. The accreted material could be
gas expelled from stars via stellar winds and supernovae, or intergalactic 
gas captured by the galaxy. As the gas accretes, the ions are heated 
by viscous dissipation and transfer their energy 
to the electrons so that the energy is radiated. However, the energy is 
inefficiently transferred from the ions to the electrons when 
the mass accretion rate is below a critical rate, and advected 
inward to the central object \citep{ny94}.  The radiative luminosity of the
Advection-Dominated Accretion Flows (ADAFs) is much less than 
that of the standard cool disk \citep{ss73}.  The ADAFs are optically thin and geometrically 
thick, i.e., quasi-spherical. The electron temperature is very hot, 
$\sim 10^9 - 10^{10}~ {\rm K}$ and relativistic. 
Angular velocity of the flow is less than the Keplerian.

In this {\it Letter} we study the stellar interactions with the ADAFs as a 
possible cooling mechanism of such a hot accretion disk. This {\it Letter} begins with 
descriptions what happens if a star meets a hot accretion disk around the SMBH in $\S$ 2. 
We calculate the heating and cooling rates per volume and present 
the volume-integrated quantities in $\S$ 3. 
We discuss observational implications and conclude in $\S$ 4.

\section{WHEN A STAR MEETS A SMBH}

Stars can be captured or disrupted by the tidal force of the SMBH in galactic nuclei, 
depending on the mass of the SMBH, the distance between the star and the SMBH, 
 a 'compactness' of the star. Since the tidal radius of the SMBH is 
proportional to $M_{SMBH}^{1/3}$, while the schwarzschild radius of the SMBH 
increases proportionally to $M_{SMBH}$, at a certain mass of the SMBH stars 
can be swallowed by the SMBH without being disrupted. For a solar type stars this
critical mass of the SMBH is about $10^8 M_{\odot}$. In other words, for 
$M_{SMBH} \ga 10^8 M_{\odot}$ a  dramatic event, such as the tidal disruption,
is expected less often.

What happens if a star passes through the accretion disk around the SMBH? Firstly,
dynamical friction causes the viscous heating. The power is given by 
$P = F_{df} \times v_{rel}$, where $F_{df}$ is the drag force, $ v_{rel}$
is the relative velocity of the star with respect to the background gas.
\citet{os99} has estimated the drag force $F_{df}$ on a star with mass $M_*$
moving through a uniform gas density $\rho$ with the relative velocity $v_{rel}$ :
\begin{equation}
F_{df}=-4 \pi I \biggl(\frac{GM_*}{v_{rel}} \biggr)^2  \rho,
\end{equation}
where the negative sign indicates that the force acts in the opposite 
direction of the star, $G$ is the gravitational constant. 
The  coefficient $I$ depends on the Mach number, 
${\cal M} \equiv v_{rel}/c_s$, where $c_s$ is the sound speed of the medium. 
In the limit of a slow moving ${\cal M} \ll 1$, $I_{subsonic} \rightarrow 
{\cal M}^3/3$, so that the resulting $F_{df}$  is proportional to the relative
speed of the star. In the limit of a fast moving ${\cal M} \gg 1$, 
$I_{supersonic} \rightarrow \ln(v_{rel}t/r_{min})$, where $r_{min}$ is 
the effective size of the regime where the gravity of the star dominates. 
Suppose that the accretion flows 
around the SMBH occur via the ADAFs. The ADAFs  have a quasi-spherical 
morphology \citep{ny95a} and the orbital velocity of the flows is sub-Keplerian. 
These two effects imply that the magnitude of the hydrodynamic drag is 
insensitive to the orientation of the stellar motion with respect to 
the rotation of the accreting gas (cf. Narayan 2000). Under such circumstances,
$v_{rel} \sim c_s \sim v_K$, where $v_K$ is the Keplerian speed, and 
therefore the Mach number is  of  order unity. We choose the supersonic 
estimate of $I$, as it gives an upper limit on the heating due to the drag force.
As we set $r_{min}$ to the gravitational capture radius, and $v_{rel}t
\rightarrow r_{max}$, where $r_{max}$ is the size of the system, $I$ is given as
\begin{equation}
I \sim \ln \biggl(\frac{r_{max}v_{rel}^2}{GM_*} \biggr).
\end{equation}
By setting $v_{rel}=v_K$, we have
\begin{equation}
I \sim \ln \biggl(\frac{M_{SMBH}}{M_*} \biggr).
\end{equation}

On the other hand, the stellar emission may heat or cool a gaseous medium, 
depending on the ambient environment. In the ADAFs a star and the stellar
motion may enhance the cooling by bremsstrahlung and Comptonization 
processes. The gas density in front of the  star may  be increased 
as the motion of the star may compress the gas. According to \citet{sg83}, 
bremsstrahlung cooling rate per volume is increased as the density 
increases, since the bremsstrahlung  cooling rate is proportional 
to the square of the density. 
Comptonization is also possible because the electrons in the ADAFs are 
relativistic. Radiation emitted by the star is an important 
source of soft photons. The spectrum of the emitted radiation can be
approximated as a blackbody. The outgoing flux at radius $R$ is
given by
\begin{equation}
F_*(R)=\frac{L_*}{4 \pi R^2},
\end{equation}
where $L_*$ is the stellar luminosity. When the star is at a distance 
from the central SMBH $d$, the distance from an arbitrary position in 
the accretion disk to the star $R$ is related 
with the distance from the position to the SMBH $r$ as given by
\begin{equation}
R^2=\Bigl|d^2 +r^2 -2rd \cos \theta \Bigr|,
\end{equation}
where $\theta$ is the angle between two position vectors ${\bf Or}$
and ${\bf Od}$ with respect to the central SMBH.
Now we are in a position to calculate the Comptonization of the stellar 
flux (cf. Narayan \& Yi 1995b). We have the stellar cooling rate as
\begin{equation}
q^{-}_{*,C}=3\frac{F_*}{R} \biggl( \frac{\theta_e}{x_b} \biggr)^3
\biggl\{\frac{\eta_1}{3}\biggl[\biggl(\frac{x_{max}}{\theta_e}\biggr)^3
-\biggl(\frac{x_c}{\theta_e}\biggr)^3 \biggr]-\frac{\eta_2}{3+\eta_3}
\biggl[\biggl(\frac{x_{max}}{\theta_e}\biggr)^{3+\eta_3}-\biggl(
\frac{x_c}{\theta_e}\biggr)^{3+\eta_3} \biggr] \biggr\},\label{eq:qc}
\end{equation}
where $\theta_e=kT_e/m_ec^2$, $k$ being the Boltzmann constant, $T_e$ being
the electron temperature, $m_e$ being the electron mass, $c$ being
the speed of light, $x_b=h\nu_b/m_ec^2$, $h$ being the Plank constant,
$\nu_b =5.61 \times 10^{10}T_*$, $x_{max}=max(x_b, 3\theta_e)$,
$x_c$ is given by the critical frequency $\nu_c$, $\eta_{k}^{'}~$s are 
 defined by \citet{der91}.  

The cooling rate of the accretion disk due to the stellar light is dependent upon 
the position of the star in the accretion flows, and its physical 
properties, such as, the mass accretion rate, the temperature, the 
density. The stellar cooling rate per volume due to Comptonization becomes 
relatively important than  those due to other processes of accretion disk cooling 
when the mass accretion rate becomes small and the star is at large distance.

\section{STELLAR COOLING OF DISK}

We adopt the following dimensionless variables : mass of 
the SMBH $m=M/M_\odot$;  radius from the SMBH $r=R/R_g$, 
where $R_g=2GM/c^2=2.95 \times 10^5~ m~{\rm cm}$; 
and mass accretion rate $\dot{m}=\dot{M}/\dot{M}_{\rm Edd}$,
where $\dot{M}_{\rm Edd}=L_{\rm Edd}/\eta_{\rm eff} c^2= 
1.39 \times 10^{18}~ m~ {\rm g~ s^{-1}}$ (the Eddington 
accretion rate assuming $\eta_{\rm eff}=0.1)$. We model the accreting
gas as a two-temperature plasma
assuming that the flows are spherically symmetric. 
The quantities of interest are the volume-integrated 
quantities which are obtained by integrating throughout 
the volume of the flows. As canonical values in a model 
for the ADAFs parameters are taken to be $r_{min}=3$, $r_{max}=10^5$,
$\alpha=0.3$, and $\beta=0.5$ (see, e.g., Narayan and Yi 1995b). 
We assume that the star in the accretion disk is a large red giant star, and
physical parameters of a  large red  giant star are quoted from \citet{all73}.

Provided that the background gas environment is described by the ADAFs model
and that stellar cooling rate per volume due to Comptonization $q^{-}_{*,C}$
is given by Eq.~\ref{eq:qc}, the volume-integrated cooling rate
$dQ^{-}_{*,C}$ over the spherical shell at $r$ can be obtained as shown in Figure 1.
We plot $dQ^{-}_{*,C}$ with other volume-integrated cooling rates 
as a function of $r$. The continuous line represents the $dQ^{-}_{*,C}$,
the dotted line and the dashed line represent volume-integrated cooling rate 
due to synchrotron cooling $dQ^{-}_{sync}+dQ^{-}_{sync,C}$ and
 bremsstrahlung cooling $dQ^{-}_{br}+dQ^{-}_{br,C}$. Note that the 
Comptonization of soft photons from synchrotron radiation and 
 bremsstrahlung radiation is included. 
Various cooling terms are computed such that two energy equations
for the ions and for the electrons are met \citep{ny95b}. In calculating the 
stellar cooling effect we ignored the bremsstrahlung cooling since
the density increase ahead of the stellar motion will result in negligible
effects as the bremsstrahlung cooling is a local effect compared 
with the stellar  Compton cooling. Dropping out the bremsstrahlung cooling 
and taking the heating rate of the supersonic case make
the total cooling estimate obtained  in this analysis  as a lower
limit of the  stellar cooling.  
As shown in Figure 1, the stellar cooling effect 
becomes important as the mass accretion rate is  smaller and 
as the star locates farther from the central SMBH. Comptonization of stellar 
photons dominates at the position of the star in the ADAFs when the mass
accretion rate is small.
In Figure 2 we show that the electron temperature in the left column
and the radio spectrum of the ADAFs in the right column. 
We show the case that the star is at 10 $r_g$, that is $r_*=10$, 
in Figure 2 as an example.
Since the dominant effects on the spectrum is due to the inner parts of 
the ADAFs, the stellar cooling at closer to the SMBH changes the spectrum 
more significantly.  
The electron temperature is averaged over the volume of the shell, and the
corresponding synchrotron spectrum of the flows is plotted next to 
the temperature plot.
The continuous line and the dotted line indicate the electron temperature 
and the radio spectrum of the ADAFs with the stellar cooling, and of the ADAFs
without the stellar cooling, respectively. 
As shown in $\dot{m}=10^{-5}$ case, Comptonization
of stellar soft photons cools out to $r\sim 10^3$, and the spectrum is
modified enough to be  observed. Since the synchrotron radio spectrum 
is a strong function of the electron temperature (see Mahadevan 1997),
the dominant effect is expected to be observed in the radio regime.
The suppression of the radio
spectrum due to stellar cooling is the greatest at the frequency corresponding
to the position where the star cools the accretion disk. It can be
understood by the fact that the synchrotron radio emission of the ADAFs at 
 each frequency is closely related to a specific radius. For instance, the 
emission at higher frequencies originates at smaller radii, 
or closer to the central supermassive black hole.

\section{DISCUSSION AND CONCLUSION}

Physical Consequences of flying-by of a star in a hot accretion disk such as
the ADAFs are the increase of the total X-ray flux due to Comptonized photon and
the decrease of the electron temperature and subsequently 
the radio flux of the hot accretion disk. The X-ray flux change due to the
stellar Comptonization is inappreciable. On the other hand, the radio flux change 
due to the change in the electron temperature of the ADAFs is sensitive enough.
 In Figure 2 we show the radio spectrum
of the ADAF model due to synchrotron emission and the suppression of
the luminosity due to the stellar Comptonization.
Of course, the conclusion depends on changes of the structure of 
the accretion disk due to the presence of a star and on responses of the star to
the strong gravity of the SMBH.  
A star interacting with a supermassive black hole becomes 
vulnerable to expansion and compression which may cause oscillating
or enhancement of nuclear reaction rate. The emitted stellar flux is an important
factor in this analysis. 
The actual cooling rate could be also higher because the bremsstrahlung cooling
may play a  role and because the heating may be less efficient
than we estimate in this analysis.
As the number of surrounding stars and 
their distribution are roughly known, the total suppression of the radio flux
due to the stellar cooling can be obtained. For instance,
suppose $N_* \sim 10^6 - 10^7~ {\rm pc^{-3}}$ in a galactic nucleus, the number
of stars in $r \sim 10^3~ {\rm r_g}$ in the accretion disk around 
$M_{SMBH} \sim 10^8 M_{\odot}$ is a few. Thus, the total
radio flux will be suppressed further by a factor of few. Note, however, that
uncertainty in estimating the total observational radio flux 
could be involved when stars accumulate on orbits between the influence radius$r_h$ 
and the core radius $r_c$, and the density of stars in the cusp bound to 
the black hole may exceed $N_*$. 

Assuming an isotropic distribution of velocities of the radius $r_h$ at which the
central SMBH dynamically influences the stellar motion, then the frequency with
which a star would pass within a distance $r$ from the SMBH can be approximated by
\begin{equation}
\approx 10^{-3}~M^{4/3}_{8} \Biggl(\frac{N_*}{10^6~ {\rm pc^{-3}}}\Biggr) 
\Biggl(\frac{\sigma}{300~ {\rm km~s^{-1}}}\Biggr) \Biggl(\frac{r}{r_t}\Biggr)~~ {\rm yr^{-1}},
\end{equation}
 where $M_8$ is the SMBH mass in units of $10^8 M_{\odot}$, 
$\sigma$ is the virial velocity of the stars, $r_t$ is the tidal radius (cf. Rees 1988).
It could be used to compare the stellar cooling event rate with 
the tidal disruption event rate. For instance, a star is tidally disrupted when 
the star passes the SMBH within the tidal radius. Therefore, the tidal disruption event 
may occur once at most every $\sim 10^{3}~  {\rm yrs}$ when the star passes the SMBH with 
$M_{SMBH}=10^8~ M_{\odot}$ at a distance of $r=r_t$.
 The stellar cooling event, however, 
may occur when a star passes the SMBH  at a distance 
farther than $r=r_t$ so that the stellar event rate could be much larger.
In the example we present in the {\it Letter}, $r_t \approx r_g$ for $M_{SMBH}=10^8~
M_{\odot}$, and the stellar cooling event rate could be larger than
the tidal disruption rate by at least a couple of orders as long as the mass
accretion rate is small. 
 
Other observational implications of the stellar cooling are as follows. Firstly,
as $dQ^{-}_{*,C}$ depends on the position of the star in the accretion disk
the frequency at which the suppression of the radio flux is the greatest varies
according to the stellar positions in the accretion disk. If one may monitor
the luminosity changes with the observed frequency one may
estimate the stellar orbit in the center of galaxies.
Secondly, one may distinguish whether there is a cool and thin accretion disk 
or there is a hot and thick accretion disk in a galaxy. The stellar cooling process
is available only  when the electrons are relativistic. Therefore, given that
such an observation is made, it could be an evidence of the presence of a hot 
accretion disk in the center of the galaxy.

\acknowledgments

The author thank I. Yi for useful discussions and C.-S. Choi for helpful comments.

\clearpage

\figcaption[fig1.ps]{
The volume integrated cooling rate over the spherical shell 
due to various mechanisms are shown as a function of $r$ in log scales
where $m=10^8$.
The continuous line represents the $dQ^{-}_{*,C}$,
the dotted line and the dashed line represent  
due to synchrotron cooling $dQ^{-}_{sync}+dQ^{-}_{sync,C}$ and
 bremsstrahlung cooling $dQ^{-}_{br}+dQ^{-}_{br,C}$.
The $dQ^{-}$'s are in $ {\rm ergs~s^{-1}}$.
The mass accretion rate and the position of star $r_*$ are shown in 
each panel. 
\label{fig1}}

\figcaption[fig2.ps]{
We show that the electron temperatures in the left column
and the radio spectra of the ADAFs in the right column in ${\rm ergs~s^{-1}}$  
in log scales.
The continuous line and the dotted line indicate the electron temperature 
and the radio spectrum of  the ADAFs with the stellar cooling, and of the ADAFs
without the stellar cooling, respectively. 
Note that the difference between the two accretion flows with and without
the stellar cooling is indistinguishable  when $\dot{m}=10^{-3}$.
\label{fig2}}

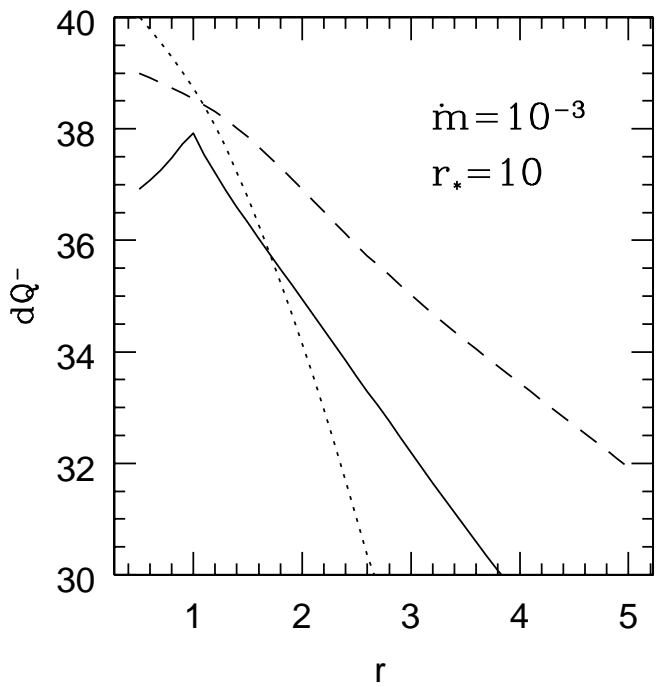
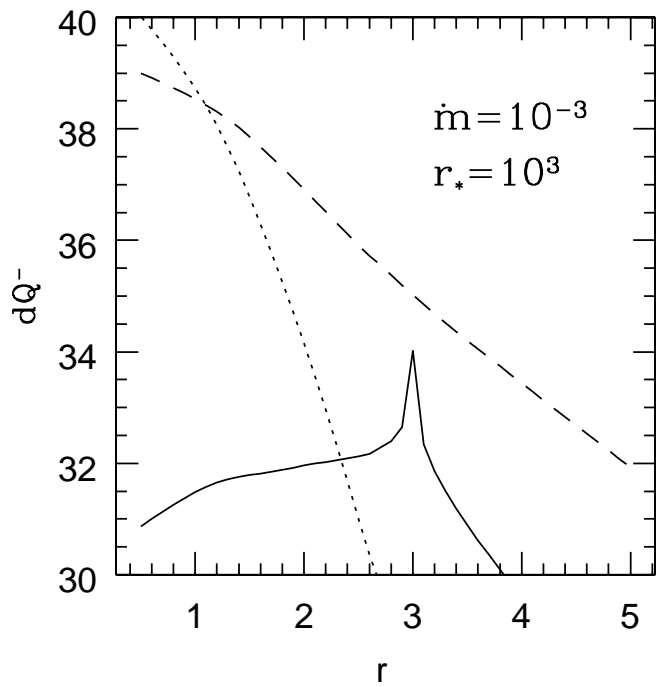
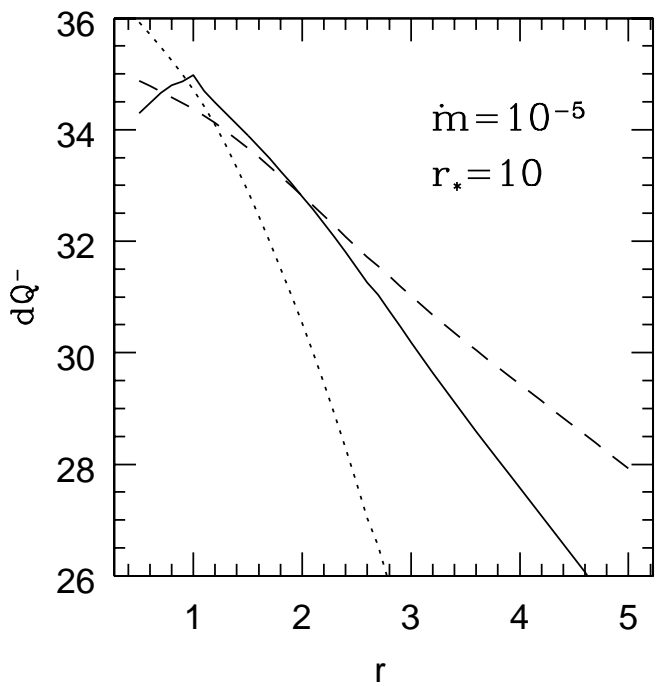
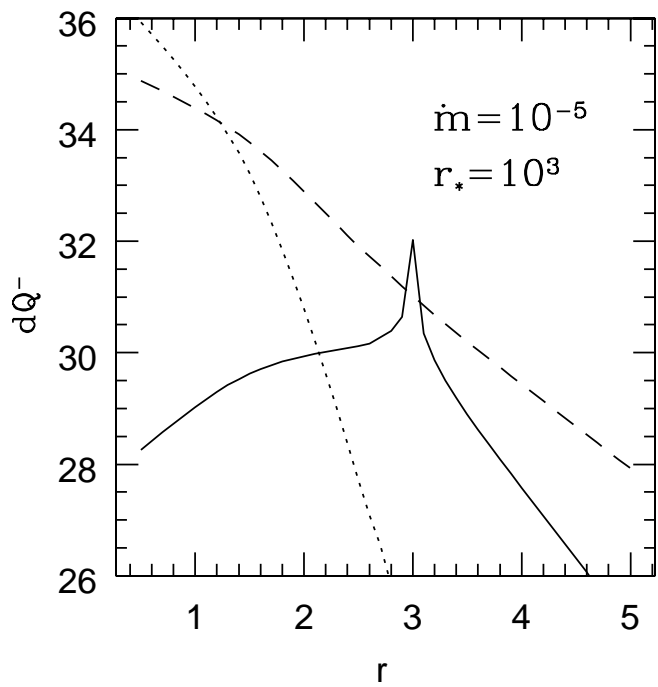
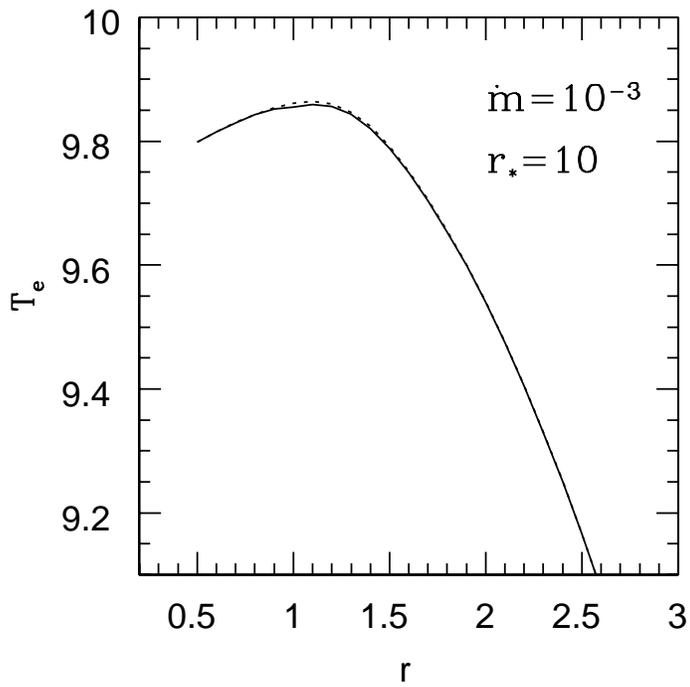
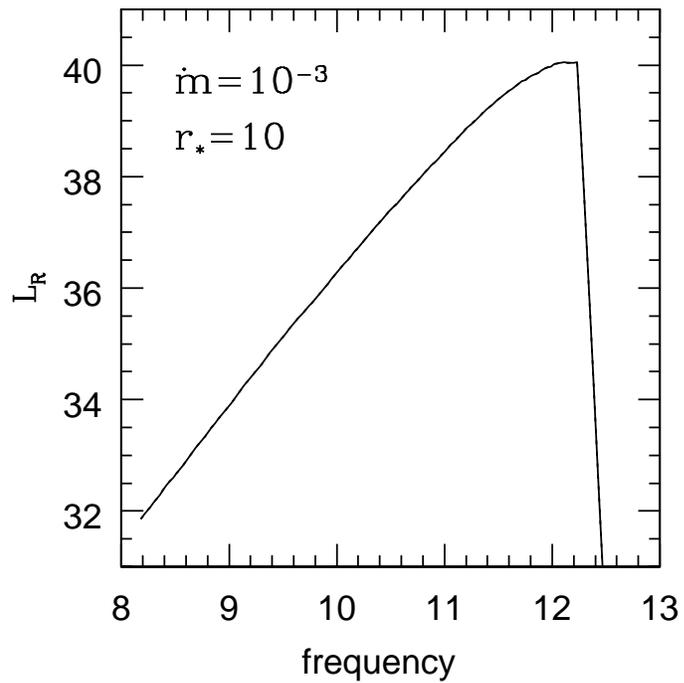
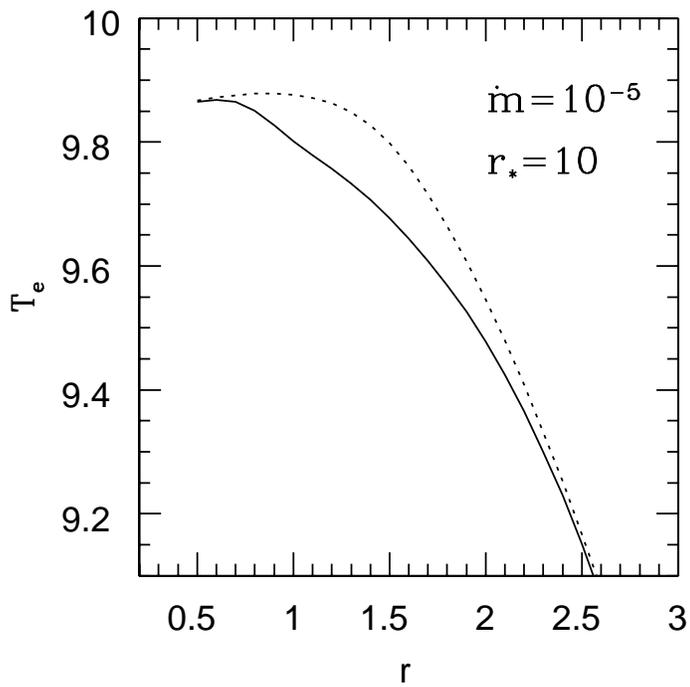
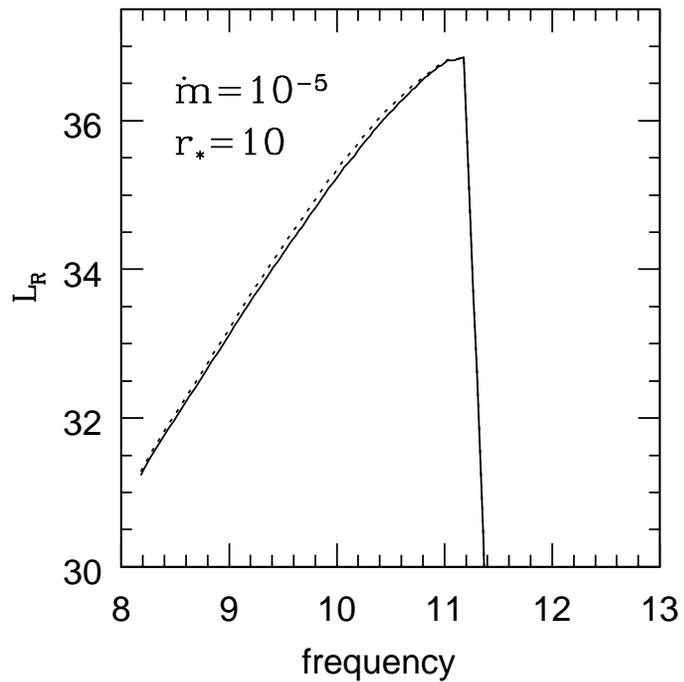
\end{document}